\documentclass[fleqn,usenatbib]{mnras}

\usepackage[T1]{fontenc}
\usepackage{ae,aecompl}

\usepackage{graphicx}	
\usepackage{amsmath}	
\usepackage{amssymb}	
\usepackage[percent]{overpic}

\title[New Avenues for Thermal Inversions in hot Jupiters]{New Avenues for Thermal Inversions in Atmospheres of hot Jupiters}

\author[Gandhi \& Madhusudhan]{
Siddharth Gandhi,$^{1}$\thanks{E-mail: sng29@ast.cam.ac.uk}
Nikku Madhusudhan,$^{1}$\thanks{E-mail: nmadhu@ast.cam.ac.uk}
\\
$^{1}$Institute of Astronomy, University of Cambridge, Madingley Road, Cambridge, CB3 0HA, UK
}

\date{Accepted XXX. Received YYY; in original form ZZZ}

\pubyear{2018}

\begin{document}
\label{firstpage}
\pagerange{\pageref{firstpage}--\pageref{lastpage}}
\maketitle

\begin{abstract}
Thermal emission spectra of hot Jupiters have led to key constraints on thermal inversions (or `stratospheres') in their atmospheres with important implications for their atmospheric processes. Canonically, thermal inversions in hot Jupiters have been suggested to be caused by species such as TiO and VO which have strong visible opacity to absorb incident starlight. We explore two new avenues for thermal inversions in hot Jupiters, exploring both the visible and infrared opacities in their atmospheres. Firstly, by exploring a range of metal-rich species we find that four species  (AlO, CaO, NaH and MgH) provide visible opacities comparable to TiO/VO and can cause strong inversions with reasonable abundances. Secondly, we show that a low infrared opacity caused by a low H$_2$O abundance, e.g. through a C/O$\sim$ 1, can also lead to strong thermal inversions even with low abundances of the visible absorbers mentioned above. We find that increasing the C/O ratio towards unity requires almost 2 orders of magnitude lower abundances for the visible absorbers in order for an inversion to form. Finally, we explore the thermal inversion in WASP-121b and find that it can be explained by all the visible absorbers listed above for different C/O ratios. Our study demonstrates the importance of both the refractory and volatile species in governing the physicochemical processes in hot Jupiter atmospheres. Spectroscopic observations in the visible have the potential to detect the newly proposed refractory species that can cause thermal inversions in addition to TiO and VO.  
\end{abstract}



\section{Introduction}
\label{introduction}

\begin{table*}
	\centering
	\caption{Summary of developments in the theory of thermal inversions in strongly irradiated giant planets.}
	\begin{tabular}{|p{3.49cm}|p{14cm}|} 
		\hline
		Study & Investigation into Thermal Inversions on hot Jupiters\\
		\hline
		\cite{hubeny_2003} & Proposed \textit{TiO} and \textit{VO} as a candidate for thermal inversions in hot Jupiters.\\
		\cite{fortney_2008} & Introduced a two-class system for classifying hot Jupiters dependant on the atmospheric temperature.\\
		\cite{spiegel_2009} & Investigated depletion of \textit{TiO} and \textit{VO} in hot Jupiter atmospheres due to gravitational settling.\\
		\cite{zahnle_2009} & Proposed \textit{HS} and \textit{S$_2$} as candidates for thermal inversions at intermediate temperatures ($\sim$1000-2000K).\\
		\cite{knutson_2010} & Investigated effect of stellar activity and concluded inversions are more likely with lower UV flux.\\
		\cite{madhu_2011} & Investigated dependence of \textit{TiO} and \textit{VO} on C/O ratio.\\
        \cite{molliere_2015} & Suggested atomic \textit{Na} and \textit{K} are capable of leading to thermal inversions for a C/O ratio near 1.\\
		\cite{beatty_2017_k13} & Proposed that higher log(g) suppresses inversions due to cold-trap processes.\\
		\cite{lothringer2018} & Proposed \textit{Fe}, \textit{Mg}, \textit{SiO} and metal hydrides as candidates for inversions on planets with $T_\mathrm{eq}$>2500 K.\\
        \textbf{This work} & Propose four species, \textit{AlO}, \textit{CaO}, \textit{NaH} and \textit{MgH} for causing thermal inversions. Investigated the effect of the infrared opacity on thermal inversions.\\
		\hline
	\end{tabular}
    \label{tab:prev_work}
\end{table*}

Atmospheric characterisation of exoplanets has entered a new era, with high precision and high resolution spectral data available for a growing number of planets \citep{Madhu_review2016}. Transiting exoplanets have been characterised using both emission and transmission spectroscopy for atmospheric studies thanks to instruments such as the Hubble WFC3 spectrograph \citep[][]{deming_2013, madhu_2014, kreidberg_2014, sing_2016}. In recent years there has also been significant development in ground-based observations, from high precision spectra in the visible \citep{sedaghati_2017} to very high resolution spectroscopy in the near infrared \citep[][]{snellen_2010,birkby_2013}. We have also been able to find molecules and investigate the temperature profiles through direct imaging of exoplanets \citep[e.g.][]{konopacky_2013, macintosh_2015}. With this we now have a wealth of observations for a range of exoplanets that encompass both strongly irradiated and non-irradiated planets over a wide range of temperatures. We have been able to detect and characterise the atmospheric temperature profiles and chemistry of exoplanets like never before and with future instruments (e.g JWST) we will be able to explore in even more detail the processes that occur on such planets.

Emission spectroscopy in particular is able to probe the temperature structure of the dayside atmosphere in detail due to its sensitivity to the thermal gradients. One of the major areas in this direction has been the study of thermal inversions in highly irradiated exoplanets, particularly hot Jupiters. Thermal inversions are regions of the atmosphere where the temperature increases with altitude, similar to the stratosphere on Earth. This is caused by entrapment of the incident stellar irradiation due to the presence of strong UV/visible absorption, such as O$_3$ for Earth's stratosphere. In hot Jupiters, molecules such as TiO and VO have been suggested to play a similar role \citep{hubeny_2003,fortney_2008}. The larger and hotter extrasolar giant planets have been thus far the most studied thanks to their strong spectral signatures. Initial observational inferences of stratospheres, or thermal inversions, were based mostly on photometric data \citep{burrows_2007,burrows_2008,knutson_2008,knutson_2009,madhu_2010}, several of which have since been revised \citep{diamond-lowe_2014}. More recently, inferences of thermal inversions have been possible with HST WFC3 observations \citep[][]{haynes_2015, evans_2017,sheppard_2017}. These planets are of great interest as only a handful of species have been suggested to be capable of leading to a thermal inversion in such atmospheres. Their strong effect on the temperature profile results in significant observable consequences in the thermal emission spectra of exoplanets. A review on thermal inversions can be seen in \cite{madhu_2014_review}.

The first theoretical work on thermal inversions on hot Jupiter atmospheres was by \cite{hubeny_2003}. Here they proposed two molecules, TiO and VO, which both had sufficiently strong absorption in the visible that they should be capable of causing thermal inversions at or near their expected solar abundances on hot Jupiters. These species are only expected to exist in the atmospheres of planets with a temperature exceeding 2000K, but a significant number of irradiated giant planets so far discovered exceed this temperature on their dayside (e.g WASP-33b, WASP-12b, WASP-18b). This work was built upon by \cite{fortney_2008}, who also modelled atmospheres with TiO and VO to calculate radiative-convective equilibrium P-T profiles of the atmospheres of several known hot Jupiters. They propose a two-class hot Jupiter classification, dependant on the equilibrium temperature being above or below $\sim$1500K. Below $\sim$1500K, planets were expected to be too cool to host TiO or VO in their gaseous state. Therefore only planets above this temperature are expected to have the TiO/VO in sufficient abundances to lead to thermal inversions. 

Subsequent studies investigated other factors that influence the possibility of thermal inversions on hot Jupiters. A detailed investigation of the effect of TiO across a number of hot Jupiters was carried out by \cite{spiegel_2009}. In this study, the vertical upwelling in the atmosphere and the day-night redistribution were also considered. They proposed that when the gravitational settling was considered, an atmospheric temperature higher than $\sim$1800K was required for TiO to be abundant enough in the observable atmosphere to cause thermal inversions. Therefore, they proposed only the hottest exoplanets were capable of thermal inversions. They also determined that VO is less likely to play a significant role in thermal inversions on hot Jupiters due to its weaker cross section and lower equilibrium solar abundance compared to TiO. 

\cite{zahnle_2009} proposed that sulphur species, such as HS and S$_2$, may be responsible for thermal inversions in the stratospheres for planets with intermediate temperatures, ranging from $\sim$1000-2000K. Both of these species possess significant UV absorption (in particular HS) and thus can lead to radiative heating rates that are significant enough to lead to thermal inversions. \cite{knutson_2010} investigated the effect of the star's UV flux, and whether spectrally active species would remain stable on such exoplanets. The most active stars with the strongest UV fluxes were the least likely to harbour thermal inversions, given that such visible absorbers would most likely be dissociated by the incident UV irradiation. Therefore they predicted that thermal inversions would be more likely on hot Jupiters if the incident UV flux was low enough for  inversion-causing species to exist. Inversions have also been suggested to be caused by thermal instabilities in the presence of ohmic dissipation in highly irradiated atmospheres from general circulation models \citep{menou_2012}. 

\cite{madhu_2011} investigated the equilibrium abundances of TiO and VO as the C/O ratio varied. They found that a C/O ratio of 1 reduced the abundance of these species by a factor of $\sim$100, thereby reducing the visible opacity of the atmosphere. \cite{madhu_2012} proposed a 2D classification scheme, depending on the atmospheric temperature and the C/O ratio for hot Jupiters. When the C/O ratio exceeds unity, the available oxygen preferentially binds to carbon to form carbon monoxide. The atmosphere is thus depleted of oxygen to form species such as H$_2$O, TiO and VO. \cite{molliere_2015} on the other hand found that a C/O ratio near 1 also resulted in thermal inversions from the gaseous atomic species Na and K. They discovered that with an atmospheric temperature over 2000K, thermal inversions were seen in the model atmospheres with F5, G5 and K5 stars at pressures of $\sim$0.1-0.01 bar. This results from the atomic line wings, which result in significant absorption of the visible flux and therefore emulate the role of TiO or VO to lead to a thermal inversion. More recently, \cite{beatty_2017} analysed the spectrum of an irradiated brown dwarf and hypothesised that the lack of a thermal inversion seen was due to the high surface gravity. The higher log(g) would likely enhance cold-trap processes which prevent TiO/VO from circulating up into the photosphere and therefore reasoned that this may prevent thermal inversions from forming on more massive exoplanets.

More recent work suggesting the thermal dissociation of H$_2$O in ultra hot Jupiters (T$\gtrsim$2500 K) has shown that thermal inversions may also form more readily on such planets \citep{parmentier_2018, lothringer2018, arcangeli2018}. The inversions are driven by the lack of H$_2$O, which causes the photospheric cooling to be more inefficient \citep{molliere_2015}. Thus thermal inversions can form due to species such as TiO and VO in the atmosphere. The dayside emission spectra from such dissociated atmospheres are expected to have weak HST WFC3 spectral features due to the weaker H$_2$O absorption but have strong CO features in the Spitzer 4.5$\mu$m bandpass as this does not readily dissociate. This mechanism of thermal dissociation has thus been proposed to explain the lack of significant WFC3 spectral features seen in several ultra hot Jupiters with an equilibrium temperature in excess of 2000K \citep[e.g.][]{kreidberg2018, parmentier_2018}. \cite{lothringer2018} have also explored equilibrium models of hot Jupiters such as KELT-9b, with T$_\mathrm{eq} \sim 4500$ K, when species such as TiO/VO begin to thermally dissociate and proposed several candidate species for inversions in the atmospheres of such extremely hot planets (see table \ref{tab:prev_work}). These species, such as Fe, Mg and SiO, remain thermally stable over a much wider range of temperatures. On the other hand, several ultra hot Jupiters do show strong H$_2$O absorption which remains a conundrum \citep{parmentier_2018}.

At the same time as equilibrium models, retrievals of hot Jupiter emission spectra has enabled atmospheric characterisation and abundance analyses from observations. Focus of research has recently moved into investigating the candidate species for thermal inversions. It has only been very recently that we have been able to obtain spectra of sufficient quality so as to infer such species in exoplanets. \cite{haynes_2015} reported signs of a thermal inversion from an emission spectrum of WASP-33b. Retrievals of WASP-33b revealed that this stratosphere was likely the effect of TiO in the atmosphere. \cite{nugroho2017} also detected TiO in WASP-33b using high resolution spectroscopy. \cite{evans_2016} obtained transmission spectra of WASP-121b with visible and near-infrared WFC3 observations and hypothesised that the spectrum was best fit with TiO or VO present in the atmosphere. This is plausible given that the equilibrium temperature for this planet is $\sim$2400K and further supported by the presence of a thermal inversion on the dayside \citep{evans_2017} found in their subsequent study. \cite{sheppard_2017} have also studied the dayside atmosphere of WASP-18b and found a thermal inversion in the infrared emission spectrum, also confirmed by \cite{arcangeli2018}. Thus there is a growing number of exoplanets which possess thermal inversions and we are now beginning to place constraints on TiO and VO through such observations. More recently, \cite{sedaghati_2017} reported a detection of TiO through the ground based transmission spectrum of the planet WASP-19b using VLT. A subsequent attempt to detect the same using GMT proved unsuccessful \citep{espinoza2019}. With new advances in instrumentation, it is likely that detections of such species becomes commonplace in exoplanets. Therefore a theoretical understanding of how thermal inversions occur in exoplanets and due to which chemical species is of paramount importance.

In this work we report three key aspects of thermal inversions on hot Jupiters. Firstly, using semi-analytic models, we explore the required atmospheric abundance for species with strong visible cross sections in order for a thermal inversion to occur in the atmosphere. These analytic models form the basis for predictions we will use and provide a good insight into the relevant radiative processes occurring in the atmosphere. We also test how well the semi-analytic prediction holds for a real atmosphere by testing against full radiative-convective equilibrium models of the atmosphere using line-by-line radiative transfer.

Secondly, we explore species beyond TiO and VO which possess a strong visible cross section and thus are capable of leading to thermal inversions. Here, we find 4 species, AlO, CaO, NaH and MgH which if present at sufficient abundance are capable of forming thermal inversions in hot Jupiters. This is done through our self-consistent radiative-convective equilibrium code GENESIS \citep{gandhi_2017}. Key to this new find has been the molecular cross sections generated from the latest and most accurate line lists available \citep{hill_2013, mckemmish_2016, patrascu_2015, yurchenko_2016, rivlin_2015, yadin_2012}.

Thirdly, we explore the effect of the infrared opacity on thermal inversions. It is well known that it is the ratio of the visible to infrared opacity that governs the formation of thermal inversions \citep{hubeny_2003,hansen_2008,guillot_2010}. Thus far, only variation of the visible opacity has been considered through species such as TiO and VO. However, we show that the infrared opacity is just as significant for causing inversions. We vary the infrared opacity through the C/O ratio of the atmosphere. The abundance of the volatile species (e.g H$_2$O, CH$_4$, NH$_3$, CO, HCN, CO$_2$, C$_2$H$_2$) is closely tied to the C/O ratio and altering this can change the abundance of these species by several orders of magnitude \citep{moses_2013}. This can thus alter the overall infrared opacity of the atmosphere and therefore the required visible opacity for an inversion. We demonstrate that with a C/O ratio near unity species such as TiO can be at sub-solar abundance and still lead to thermal inversions. We follow on to model WASP-121b, a hot Jupiter known to possess a thermal inversion \citep[][]{evans_2017}. We explore the required abundance of each of the metallic species to cause a thermal inversion in its atmosphere for both C/O=0.5 (solar) and C/O=1 for the planet. 

In what follows, section \ref{sec:theory} introduces the theory and the conditions required for a thermal inversion in an atmosphere. Section \ref{sec:vis opac} describes the effect of the visible opacity and the new species which can lead to stratospheres in these atmospheres. Section \ref{sec:c/o} discusses the importance of the infrared opacity in determining the conditions for a thermal inversion through the effect of the C/O ratio. We explore the thermal inversion in a known system, WASP-121b, in section \ref{sec:Real system} with comparisons between models with various inversion-causing species against the observations. Finally, the conclusion and implications of the work are discussed in section \ref{sec:conclusion}.

\section{Theory of Thermal Inversions}\label{sec:theory}

Thermal inversions or `stratospheres' on exoplanets require specific atmospheric conditions, but leave strong signatures on the emergent spectra. The spectral signatures allow us to study stratospheres from observations of exoplanet spectra and explore the chemical species which may be responsible. In this section we will explore the theory of thermal inversions, from prominent species which can cause inversions to the influence of volatile species such as H$_2$O. On Earth, the thermal inversion is achieved through ozone, which absorbs the strong UV radiation incident from the Sun, thus heating the upper layers of the atmosphere. Hot Jupiters can exhibit thermal inversions with quite different chemistry thanks to their high temperature and extreme irradiation. In both cases however, the method through which thermal inversions are achieved is the same, strong absorption of the incident flux at shorter wavelengths of light, typically UV/visible ($\sim$0.4-1$\micron$). The flux from the star is greatest in the visible part of the spectrum, and any absorber with a strong visible cross section that is present in significant quantities will heat the upper atmosphere. This thermal inversion effects the thermal emission spectrum received from the dayside atmosphere. Outlined below is the effect of these stratospheres on the observed spectra. We then explore semi-analytic and full line-by-line radiative-convective equilibrium models and their importance in predicting and modelling inversions on exoplanet atmospheres.

\subsection{Signatures of Thermal Inversions in Emission Spectra}
\label{sec:effect on spectrum}
Despite the relatively modest spectral coverage, recent observations have been able to reveal stratospheres in a number of hot Jupiters \citep[][]{haynes_2015, evans_2017,sheppard_2017}. These dayside emission spectra are well suited to characterise the pressure-temperature (P-T) profile, and in particular determine whether the atmosphere possesses a thermal inversion as the spectra are very strongly dependant on the thermal gradient in the photosphere. The strong influence on the observed spectrum means that through instruments such as the HST WFC3 spectrograph, tight constraints have been obtained for the temperature gradient in the photosphere.

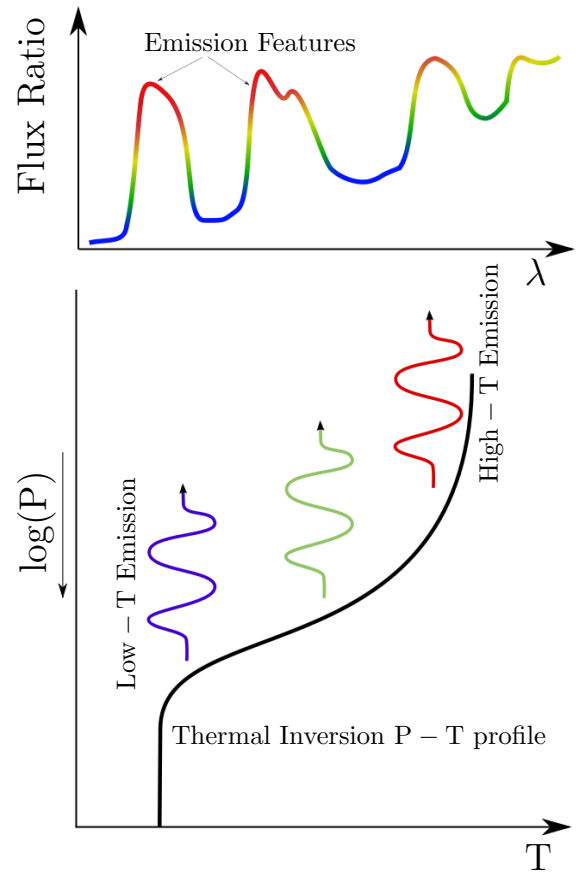
\begin{figure}
\begin{overpic}[width=\columnwidth]{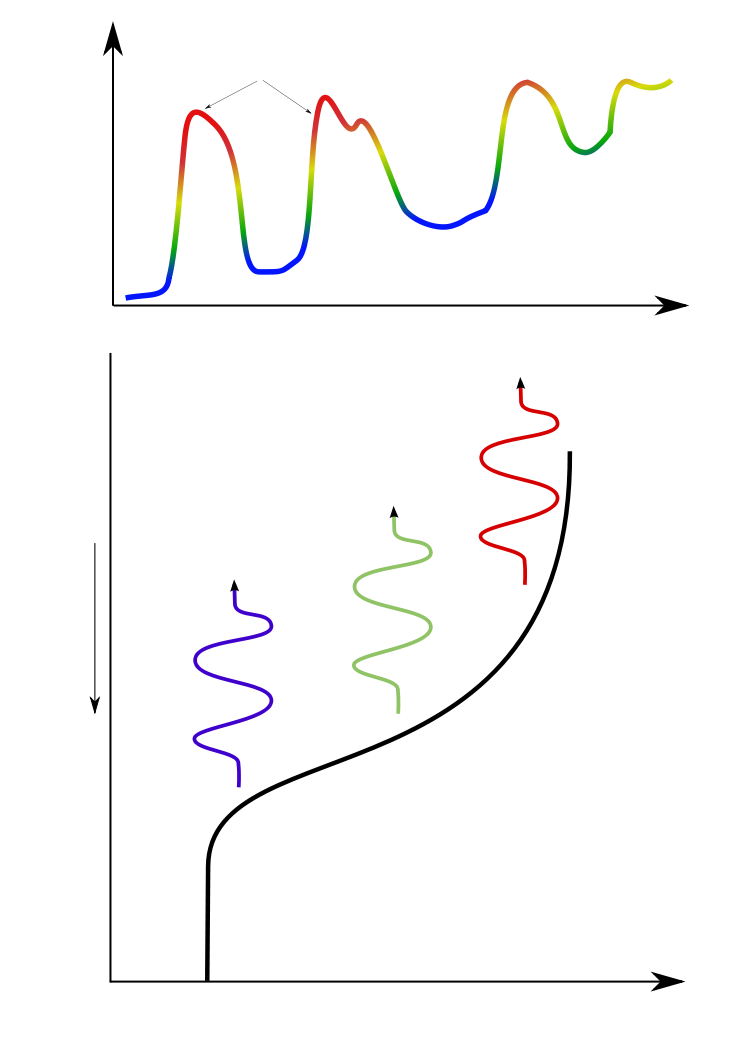}
 \put (60,2) {\LARGE $\mathrm{T}$}
 \put (60,67) {\LARGE $\lambda$}
 \put (21,16) {\large $\mathrm{Thermal \, \, Inversion \, \, P-T \, \, profile}$}
 \put (18,93) {\large $\mathrm{Emission \, \, Features}$}
 \put (15,22) {\rotatebox{90}{\large $\mathrm{Low - T \, \, Emission}$}}
 \put (55,45) {\rotatebox{90}{\large $\mathrm{High - T \, \, Emission}$}}
 \put (4,77) {\rotatebox{90}{\LARGE $\mathrm{Flux \, \, Ratio}$}}
 \put (4,35) {\rotatebox{90}{\LARGE $\mathrm{log(P)}$}}
 \end{overpic}
    \caption{Schematic showing the effect of a thermal inversion on the observed planet/star flux ratio. Lower regions in the atmosphere, where the temperature is cooler, are probed in opacity windows, i.e. wavelengths with low opacity. At wavelengths where molecular absorption is stronger, emission features occur due to the flux being emitted from higher up in the atmosphere where temperature is higher in the case of a thermal inversion (see section \ref{sec:effect on spectrum}).}
    \label{fig:schematic}
\end{figure}

Thermal inversions leave a strong signature on the observed emission spectrum of the dayside of an exoplanet \citep{fortney_2008,burrows_2008,madhu_2010}. Consider a planet who's temperature is an isotherm. In this case, regardless of the location in the atmosphere from which the emission occurs, the observed spectrum will be that of a black body at the isothermal temperature. However, if a temperature gradient is present such that the temperature increases going outwards in the atmosphere, emission features are formed in the emergent spectrum. Some wavelengths will have greater opacity than others, given that the molecular cross-sections of the constituent chemical species can vary by many orders of magnitude over the spectral range. Hence emission occurs from different points in the atmosphere. Wavelengths with low opacity will have flux emanating from lower down in the atmosphere, where the temperature is cooler. On the contrary, wavelengths with greater opacity will only have emission occurring from the uppermost layers of the atmosphere. There is a greater flux emitted in the regions of higher opacity, given that the temperature is higher here (see fig. \ref{fig:schematic}). Therefore the observed spectrum will have distinct emission features resulting from the higher opacity wavelengths. Thus \emph{emission features} result from the presence of a thermal inversion. The strength of these features depend on the abundance of the spectrally active species that causes them in the emergent spectrum, and thus we can also constrain the molecular mixing ratio of such spectrally active species. Molecular features in the emergent spectra are often broad and can cover a significant wavelength range (see fig. \ref{fig:cross_secs}), therefore can be constrained even with relatively low resolution observations \citep[][]{kreidberg_2014, line_2016}.

\subsection{Analytic Semi-Grey Model}
\label{sec:grey}

Semi-analytic models of atmospheres can provide important insights into the atmospheric processes that occur. In particular they can help understand the underlying principles of thermal inversions in what are often complex atmospheres, with a number of chemical species each with its own frequency dependant spectral features. To begin, we will consider the semi-grey analytic model by \cite{guillot_2010}. This allows us to extract the dominant physics that determines thermal inversions on exoplanets. From then, we can consider self-consistent equilibrium models to complete the picture and compare to the simple analytic model. 

The semi-grey temperature profile at an optical depth $\tau$, for a planet with internal temperature $\mathrm{T_{int}}$ and irradiation temperature $\mathrm{T_{irr}}$ is given by \citep{guillot_2010},
\begin{align}
T^4 = \frac{3}{4}\mathrm{T_{int}}^4 (\frac{2}{3} + \tau) + \frac{3}{4}\mathrm{T_{irr}}^4 \mu(\frac{2}{3} + \frac{\mu}{\gamma} + (\frac{\gamma}{3\mu} - \frac{\mu}{\gamma})e^{-\gamma \tau/\mu}),
\end{align}
where $\gamma = \kappa_\mathrm{vis}/\kappa_\mathrm{ir}$ denotes the ratio of visible to infrared opacity. Here, $\tau$ refers to the monochromatic optical depth of the atmosphere. $\mu^2 = K_\mathrm{vis}/J_\mathrm{vis}$ is defined by the ratio of moments of the specific intensity in the visible $I_\mathrm{vis,\mu}$,

\begin{align}
J_\mathrm{vis} &= \frac{1}{2}\int_{-1}^{1} I_\mathrm{vis,\mu} d\mu,\\
K_\mathrm{vis} &= \frac{1}{2}\int_{-1}^{1} \mu^2 I_\mathrm{vis,\mu} d\mu.
\end{align}
For strongly irradiated planets, we can assume that the internal temperature is negligible compared to the irradiation temperature, $\mathrm{T_{int}} \ll \mathrm{T_{irr}}$. For there to be an inversion in the atmosphere, the temperature at small optical depth must exceed that at high optical depth,
\begin{align}
f \equiv \frac{T(\tau=0)^4}{T(\tau \gg1)^4} \approx \frac{\frac{2}{3} + \frac{\gamma}{3\mu}}{\frac{2}{3} + \frac{\mu}{\gamma}} > 1.
\end{align}
The denominator is determined at a high optical depth ($\tau >> 1$) but also below the penetration depth ($\tau < \tau_\mathrm{pen}$), where the internal heat flux of the planet dominates over the incident radiation flux. This typically occurs at $P_\mathrm{pen} \gtrsim 100-10^3$ bar \citep{gandhi_2017}. Rearranging gives us a quadratic in $\gamma$. Hence the condition on the atmospheric opacity ratio $\gamma$ is
\begin{align}
\frac{\gamma^2}{3\mu}  &- \frac{2}{3}(f-1)\gamma - \mu = 0,\\
\gamma &= \kappa_\mathrm{vis}/\kappa_\mathrm{ir} = \mu(f-1) + \mu ~\sqrt[]{(f-1)^2 +3}.\label{eqn:kv_kir}
\end{align}
Thus for a thermal inversion to take place (i.e $f>1$), $\gamma>\mu ~\sqrt[]{3}$. Under the assumption of isotropic incoming and outgoing radiation, where $\mu \approx 1/\sqrt[]{3}$ \citep{hubeny_book}, this becomes 
\begin{align}\label{eqn:gamma}
\gamma = \kappa_\mathrm{vis}/\kappa_\mathrm{ir} \gtrsim 1. 
\end{align}
Hence the visible opacity must exceed the infrared opacity in order for a thermal inversion to occur in the atmosphere. Higher infrared opacity thus requires a greater visible opacity in order to cause an inversion. The assumptions that have gone into this analytic model should be emphasised here, namely that the incident stellar flux is negligible in the infrared and that the planet's own emission occurs primarily in the infrared. The semi-grey model also assumes an outgoing and an incident beam of radiation, the two-stream approximation, and that the opacity is not a function of wavelength in the visible and the infrared. However, these results do match well with full radiative-convective equilibrium models with layer-by-layer opacity calculations as discussed in section \ref{sec:GENESIS tio} below.

\subsection{Self-Consistent Models}
\label{sec:GENESIS tio}

\begin{figure}
	\includegraphics[width=\columnwidth]{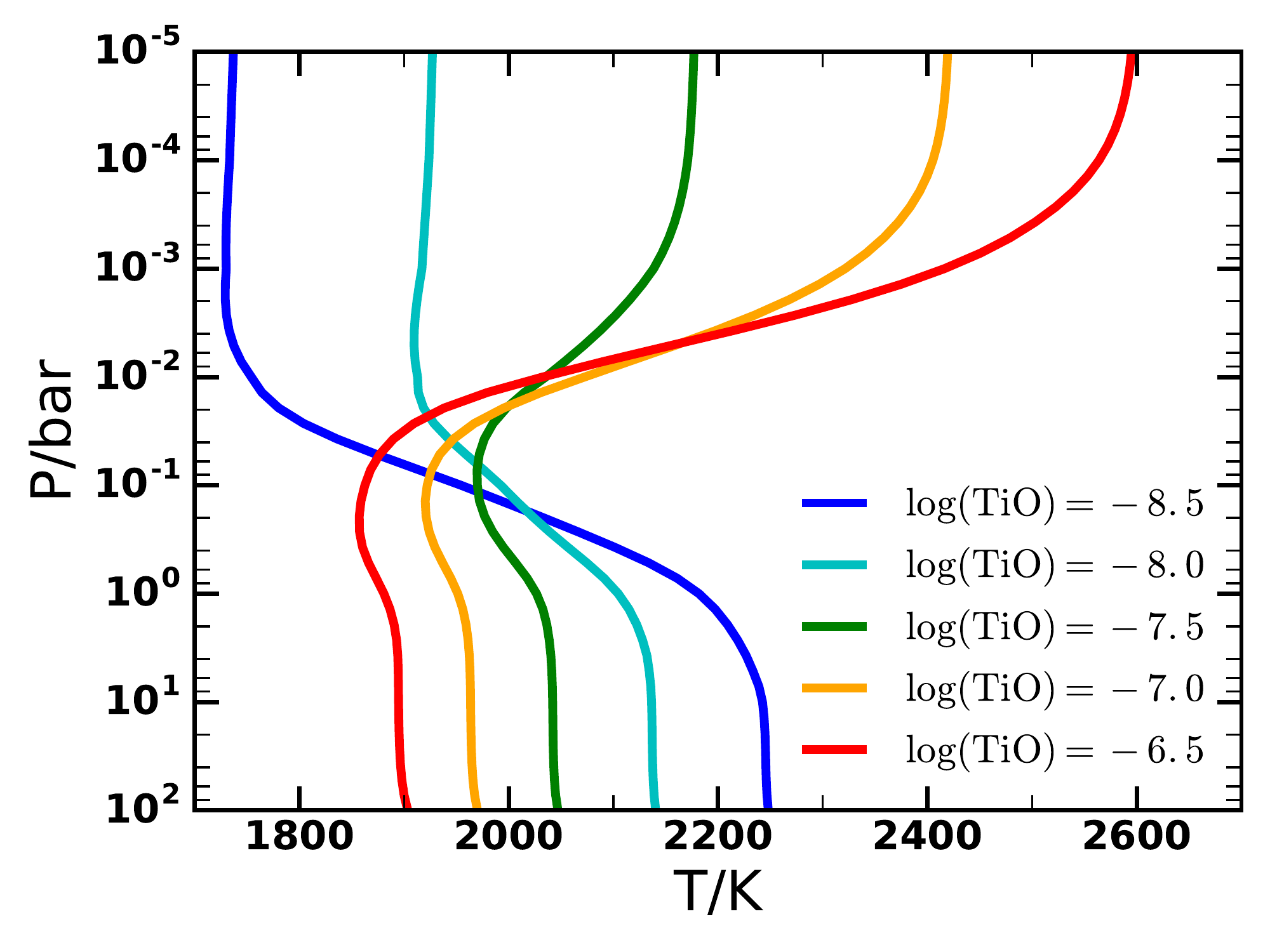}
    \caption{Effect of varying the TiO abundance on the equilibrium P-T profile of a hot Jupiter.  The P-T profiles are generated with the GENESIS model of a hot Jupiter with equilibrium temperature of 2000 K and solar equilibrium abundances for the volatile species. In this model we assume TiO is the sole source of the visible opacity.}
    \label{fig:tio}
\end{figure}

The semi-analytic grey model will now be compared against our self-consistent atmospheric model GENESIS \citep{gandhi_2017}. This model includes full line-by-line frequency dependent radiative transfer with radiative-convective equilibrium. We will test how well the condition for a thermal inversion from the semi-grey model holds given a more complex atmosphere. The algorithm used to run the equilibrium models uses the Feautrier radiative transfer method with Rybicki's linearisation scheme for temperature correction, the full details of which can be found in \cite{gandhi_2017} and \cite{hubeny_2017}. This model calculates the radiative-convective equilibrium atmospheric profile which we can compare against the semi-grey model to determine how much the frequency dependant molecular opacity affects the P-T profile. For this work we have assumed the bottom boundary condition to be the diffusion approximation. We investigated pressures of 100, 10$^3$ and 10$^4$ bar for the bottom of the atmosphere, but found no significant differences in the resultant equilibrium P-T profile and emergent spectrum. This is because these pressures are significantly below the photosphere and the internal heat flux from the planet is significantly weaker than ambient flux due to irradiation.

For illustration, we consider a hot Jupiter around a Sun-like star, with an equilibrium temperature of 2000K and a cloud-free atmosphere. We assume the infrared opacity to be contributed by the seven prominent volatile species, H$_2$O, CH$_4$, NH$_3$, CO, CO$_2$, HCN and C$_2$H$_2$ which are considered to be in chemical equilibrium \citep[][]{madhu_2012,heng_2016,gandhi_2017}. The visible opacity is assumed to arise from TiO only. The choice of this is two-fold: TiO has been observed in the visible under transmission spectroscopy \citep{sedaghati_2017} and has been suggested as a candidate for thermal inversion in the atmosphere of hot Jupiters because of its very strong visible cross-section \citep{hubeny_2003,fortney_2008}. The TiO is assumed to be uniformly mixed throughout the atmosphere and the atmospheric volume mixing fraction is varied to alter the visible opacity as needed. 

For the semi-grey model, we first need to obtain an estimate of the infrared and visible opacity of the atmosphere, $\kappa_\mathrm{ir}$ and $\kappa_\mathrm{vis}$. This is done by summing the molecular cross-sections of each of the chemical species, weighted by their mixing fractions, and then integrating over frequency following \cite{hubeny_2003}. The opacity from a species $i$ is given by
\begin{align}
\kappa_{\nu,i} &= \frac{P}{k_bT} {\sigma}_{i}(\nu),
\end{align}
where $\sigma_{i}(\nu)$ is the molecular cross section as a function of frequency $\nu$ and $k_b$ is the Boltzmann constant. The molecular cross sections are computed over a broad spectral range between 0.4-50$\mu$m. A representative temperature and pressure ($T_p = 2000$ K and $P = 0.1$ bar) were chosen to match the conditions of the photosphere. The overall opacity in the infrared is estimated through the Planck mean opacity which for a species $i$ is given by 
\begin{align}
\kappa_\mathrm{IR,i} &= \frac{\int_0^\infty \kappa_{\nu,i} B_\nu(T_p)\rm{d}\nu }{\int_0^\infty B_\nu(T_p)\rm{d}\nu}.
\end{align}
Here, the Planck function effectively weights the opacity $\kappa_{\nu,i}$ according to the emergent radiation to determine the overall infrared opacity. To determine the visible opacity for a species $j$ we use the absorption mean opacity for a solar type star ($T_s$ = 5700 K),
\begin{align}
\kappa_\mathrm{vis,j} &= \frac{\int_0^\infty \kappa_{\nu,j} B_\nu(T_s)\rm{d}\nu }{\int_0^\infty B_\nu(T_s)\rm{d}\nu}.
\end{align}
The absorption mean weights the opacity with the stellar Planck function to determine the effective absorption from species with strong visible cross-sections which absorb the incident star light which peaks in the visible.  A further discussion of the absorption mean and Planck mean opacities can be found in \cite{hubeny_2003}. By equating $\kappa_\mathrm{ir}$ to $\kappa_\mathrm{vis}$, we find the minimum required mixing fraction of TiO for which an inversion occurs as  
\begin{align}
X_\mathrm{TiO} &= \frac{\sum_i \kappa_{\mathrm{IR},i} X_i}{\kappa_\mathrm{vis,TiO}}\label{eqn:inv_abundance}.
\end{align}

We calculate the mixing fractions $X_i$ for each of the 7 volatile species at 2000K and 0.1 bar pressure in chemical equilibrium assuming solar abundances. The $\kappa_\mathrm{vis,TiO}$ term denotes the absorption mean opacity of TiO weighted with a black body at $T_s=5700$ K. We then find $\mathrm{log(X_{TiO})} \approx -7.0$ for our model atmosphere. For a general species $j$ with strong absorption in the visible, the abundance required for a certain P-T profile with a fixed $\gamma \equiv \kappa_\mathrm{vis}/\kappa_\mathrm{ir}$ is given by
\begin{align}
X_\mathrm{vis,j} &= \gamma \frac{\sum_i \kappa_{\mathrm{IR},i} X_i}{\kappa_\mathrm{vis,j}}. \label{eqn:x_vis}
\end{align}

Figure~\ref{fig:tio} shows the effect on the P-T profile of varying TiO throughout the atmosphere in the self-consistent equilibrium model. The abundance is increased from $\mathrm{log(X_{TiO})} =-8.5$ to $-6.5$, keeping the chemical composition of the volatile species fixed at solar composition. The figure shows the P-T profile inverting as expected for a TiO abundance remarkably close to $\mathrm{log(X_{TiO})} = -7.0$. This confirms that the semi-grey model gives a good estimate of the opacity required for an inversion. We have also tested this against the required abundance for a thermal inversion from the semi-analytic model for a number of species and have also found good agreement with GENESIS. Hence, we use the semi-grey model to provide a good guide to estimate the required abundance of species in order for thermal inversions to occur (see table \ref{tab:metal_inv}).

Figure \ref{fig:tio} shows that the temperature at pressures $\lesssim0.1$ bar increases as the abundance of TiO increases from $\mathrm{log(X_{TiO})} =-8.5$ to $-6.5$. This is due to increased absorption of the incident radiation in the upper atmosphere. Consequently, a smaller fraction of the stellar flux penetrates down into the deeper layers of the atmosphere, hence why the atmosphere at P$\gtrsim1$ bar is cooler as the TiO abundance is increased. Eventually at a TiO abundance of -6.5, the P-T profile is dominated by the strong visible opacity, and $\approx700$K inversion is seen between the top and bottom of the atmosphere. Throughout all of the variation of TiO abundance however, the region of the atmosphere near P$\sim0.1$ bar remains at a similar temperature. This is the photosphere where we much of the emission occurs. Therefore, the temperature gradient in the photosphere is strongly affected by the abundance of the TiO, even though the temperature here may not vary too significantly at 0.1 bar.

The solar abundance of Ti is $\mathrm{log(Ti_{solar})} \approx -7.1$, similar to that required for TiO to result in a thermal inversion. If we assume that most of the Ti will be in the form of its oxide, then the atmospheric abundance is sufficient at solar abundance to result in an inversion. However, we observe a spread in hot Jupiters with thermal inversions, which may be due to the differences in the chemical compositions of these planets. Other factors such as cold-trap process \citep{beatty_2017_k13} may also prevent thermal inversions occurring. Rain-out of TiO at temperatures below $\sim2000$K is also another important consideration \citep{spiegel_2009}. TiO will only be in its gaseous form on the very hottest of exoplanets, and may be a contributing factor as to why some of the cooler hot Jupiters such as WASP-43b \citep{kreidberg_2014} and HD209458b \citep{line_2016} show non-inverted atmospheric temperature profiles. 

\section{Effect of Visible Opacity on Inversions}
\label{sec:vis opac}

\begin{figure}
	\includegraphics[width=\columnwidth]{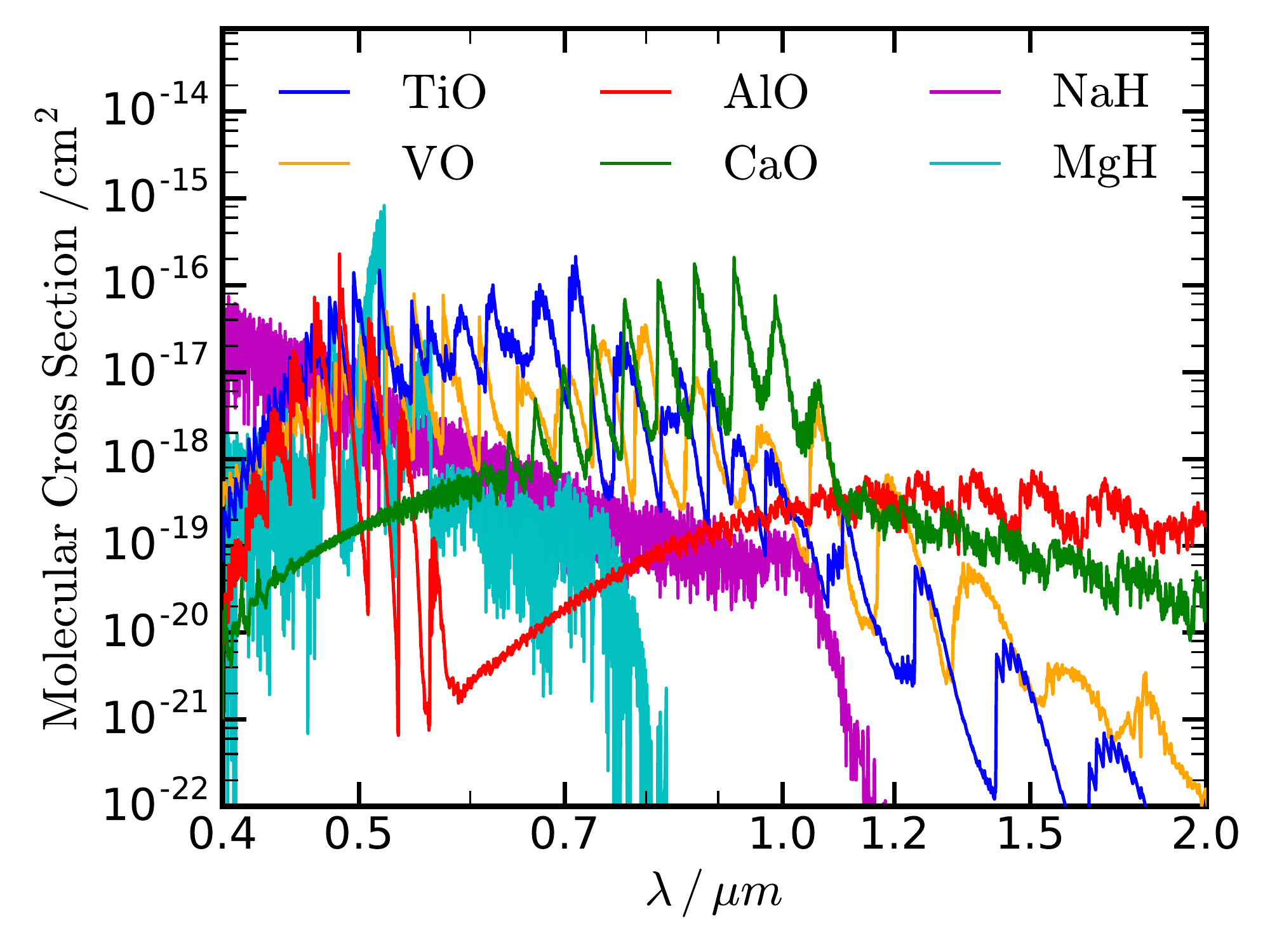}
    \caption{Molecular cross sections of chemical species with strong absorption cross sections in the visible. The cross sections were calculated at representative conditions of 0.1 bar pressure and 2000K.}
    \label{fig:cross_secs}
\end{figure}

\begin{table}
	\centering
	\caption{Table showing the solar atomic log mixing ratio and the required molecular log mixing ratios of each metallic species in order to form a thermal inversion for various C/O ratios. The solar atomic abundances were obtained from \citet{asplund_2009}. The required volume mixing ratios were determined from eqn. \ref{eqn:x_vis} and thermochemical equilibrium of the volatile infrared species at 2000K and 0.1 bar. In each model the species with strong visible opacity was included by itself with no other visible absorbers present.}
	\begin{tabular}{l|cccc} 
		\hline
        & & \multicolumn{3}{c}{Required log(Vol. Mixing Ratio)}\\
        \cline{3-5}\\
		Species & Solar &C/O = 0.5 & C/O = 1 & C/O = 1.5\\
		\hline
		TiO & -7.1 & -7.0 & -9.2 & -7.5\\
		VO &  -8.0 & -6.5 & -8.7 & -7.1\\
		AlO & -5.6 & -5.9 & -8.2 & -6.5\\
		CaO & -5.7 & -6.5 & -8.8 & -7.1\\
		NaH & -5.8 & -6.2 & -8.5 & -6.9\\
		MgH & -4.5 & -6.6 & -8.9 & -7.2\\
		\hline
	\end{tabular}
    \label{tab:metal_inv}
\end{table}

We will now investigate species other than the commonly hypothesised TiO and VO which may lead to stratospheres. Fig. \ref{fig:cross_secs} shows the molecular cross sections at 2000K and 0.1 bar for a number of metallic species generated with the method of \cite{gandhi_2017}. These cross sections were computed using the line lists obtained from EXOMOL \citep{tennyson2016} with the pressure broadening coefficients derived using the method in \cite{sharp_2007}. Fig. \ref{fig:cross_secs} shows the species which have strong visible cross sections and are therefore able to affect the equilibrium P-T profile of the atmosphere even at trace mixing ratios. As we have already seen in the previous sections, TiO at solar abundance is sufficient for a thermal inversion, even lower than this amount is sufficient in some cases (see section \ref{sec:c/o}). Therefore these species with strong visible opacity are important in determining the P-T profile and hence the emergent emission spectrum from the dayside atmosphere.

VO is another candidate often considered for a thermal inversion \citep{fortney_2008, spiegel_2009}. Like TiO, VO is predicted to only be gaseous in hot Juptier atmospheres at temperatures exceeding $\sim$1800K. Table \ref{tab:metal_inv} shows the required abundances for VO to result in a thermal inversion and the solar abundance of V \citep{asplund_2009}. Only slightly super-solar abundance of VO is sufficient to satisfy equation \ref{eqn:gamma} and thus result in stratosphere. However, any species with strong visible absorption is capable of causing thermal inversions on hot Jupiters if their abundance is significant enough. By carefully studying the species from their molecular line lists and cross sections, we are able to explore several new candidates for thermal inversions. In what follows, we determine the abundances required for each of these species to lead to an inversion for a range of model atmospheres below.

\subsection{Other Sources of Visible Opacity}

Strong molecular opacity can arise from a number of different species. As well as the well explored TiO and VO, AlO, CaO, NaH and MgH all possess strong molecular spectral features in the visible region of the spectrum as shown in fig. \ref{fig:cross_secs}. These cross sections have been derived from the latest line lists of each of these species \citep{patrascu_2015, yurchenko_2016, rivlin_2015, yadin_2012}. They all have distinct cross sections, but over the visible spectrum they provide similar strong opacity. AlO has a peak cross section near $\sim$0.5$\micron$, and CaO has a peak nearer 1$\micron$. Hydride species such as NaH and MgH also have strong enough molecular cross sections to satisfy equation \ref{eqn:gamma} at nearly solar abundance \citep[see e.g][]{lothringer2018}. We will now model the effect of each of these new species on the model atmosphere. As previously we will assume a planet with an equilibrium temperature of 2000K and equilibrium volatile chemistry.

\subsection{Effects on Equilibrium Profile}

\begin{figure*}
	\includegraphics[width=\textwidth]{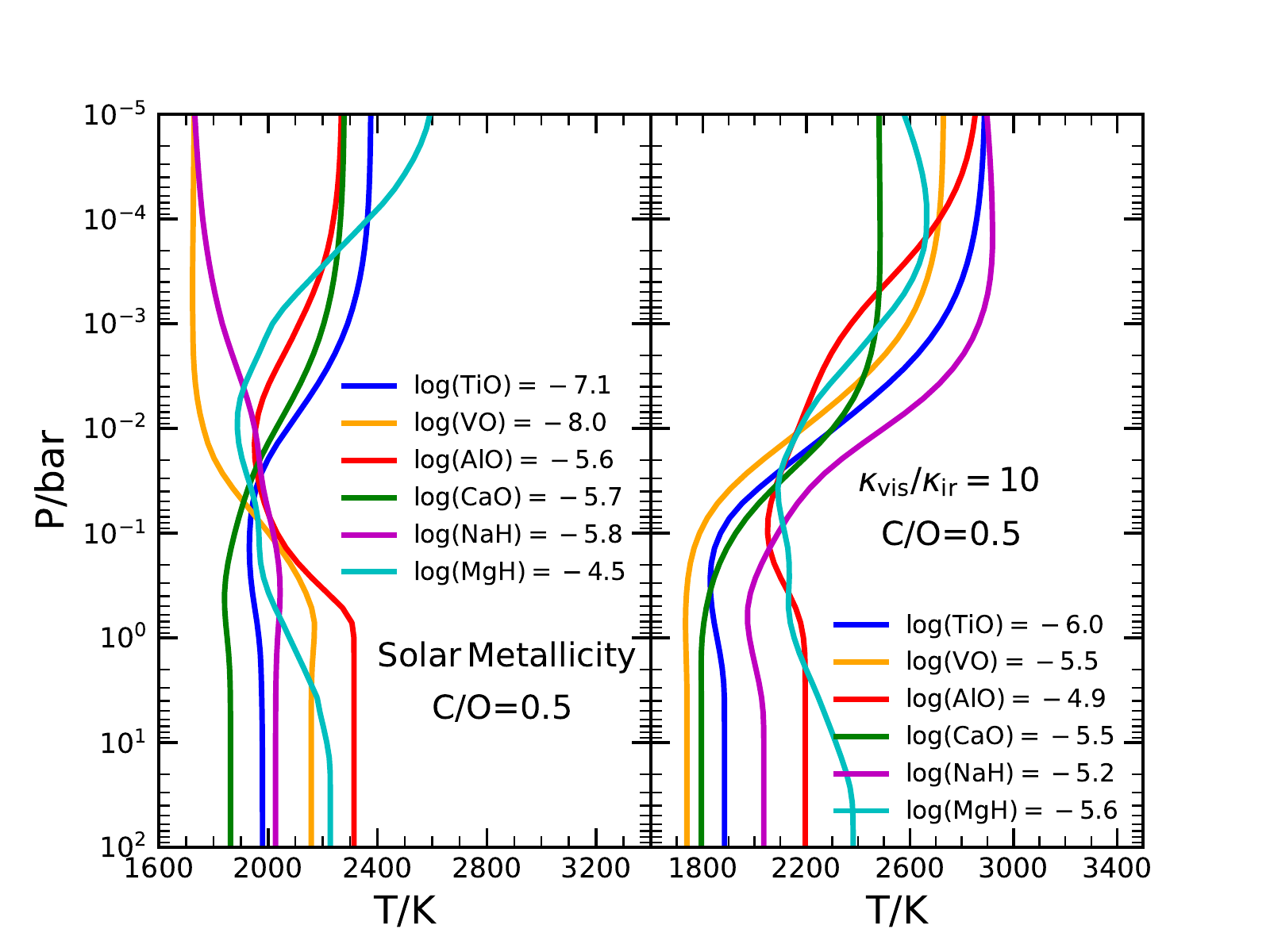}
    \caption{ Radiative equilibrium P-T profiles of model atmospheres with varying abundances of metallic species. The model assumes a hot Jupiter with an equilibrium temperature of 2000K and a solar C/O ratio of 0.5. The left panel shows the effect of solar mixing fractions for the inversion-causing refractory species. In the right panel, the abundances of the inversion-causing species are such that the ratio of visible to infrared opacity was 10. The volatile species were assumed to be in chemical equilibrium with solar abundances, and in each model the species was included in the atmosphere by itself with no other visible absorbers.}
    \label{fig:double_plots}
\end{figure*}

We determined the required abundance for each of the prominent species capable of causing thermal inversions. This was done by first calculating the absorption mean opacities for the known metallic species with strong visible cross-sections to determine $\kappa_\mathrm{vis}$. We calculated the infrared opacity assuming equilibrium chemistry for the 7 volatile species. We then equated the infrared and visible opacity to determine the required mixing fraction for an inversion, analogously with TiO in eqn. \ref{eqn:inv_abundance}. Table \ref{tab:metal_inv} shows the required abundances of each species to satisfy $\gamma =1$ from the semi-analytic model as well as their solar atomic abundances. If we assume that all of the atomic species are present in the form of their relevant molecule ( i.e., that the atom is bound only to the relevant molecule), all four of the new species, AlO, CaO, NaH and MgH are at or near the abundance required for an inversion at solar abundance. This is because despite the overall lower mean molecular cross sections compared to TiO and VO, all of their respective atomic abundances are higher than Ti or V. 

Fig. \ref{fig:double_plots} shows the effect of the metallic species on the equilibrium P-T profile of the atmosphere. The models were run with our GENESIS self-consistent radiative-convective equilibrium model, assuming equilibrium chemical compositions for the volatile species at a C/O ratio of 0.5 (solar value). As seen in the previous section, TiO results in an inversion at 0.1 bar of $\sim$500K between the top and bottom of the atmosphere at solar composition (left panel fig. \ref{fig:double_plots}). At solar abundance, VO is not present in sufficient quantities to lead to a thermal inversion, as per the prediction from the semi-grey models (see table \ref{tab:metal_inv}).

Of the new species at solar composition, AlO is at just the abundance required for an inversion and thus we can see the onset of the temperature inversion at pressures lower than 10mbar on the left panel in figure \ref{fig:double_plots}. CaO shows a similar trend to TiO, with a similar P-T profile for the atmosphere and an inversion of $\sim$400K. The similar P-T profiles for these two species is likely due to the similar molecular cross sections, which show broad coverage over all of the visible spectrum (see fig. \ref{fig:cross_secs}). NaH only shows very weak signs of a thermal inversion at 1 bar, and the temperature profile is monotonically decreasing with height. The molecular abundance at solar composition is not quite sufficient to produce a significant inversion as it is very near the critical abundance required. MgH on the other hand does show a strong molecular cross section only up to 0.8$\micron$ in fig. \ref{fig:cross_secs} and therefore displays a significantly different P-T profile to TiO, VO or CaO. MgH thus allow some of the stellar flux to penetrate down to the deeper layers due to its lack of strong opacity over all of the peak stellar flux range.

The right panel of fig. \ref{fig:double_plots} shows the effect of altering the mixing fraction of each of the metallic species through equation \ref{eqn:x_vis}. The abundance of each of the species was adjusted accordingly to maintain $\gamma \equiv \kappa_\mathrm{vis}/\kappa_\mathrm{ir} = 10$ and therefore result in a thermal inversion of similar extent for all of the species. We now see similar equilibrium P-T profiles from all of the species, however, the non-grey nature of the opacity results in some differences. AlO and MgH show the inversion occurring higher in the atmosphere (P $\sim10^{-2}$ bar) and the deepest layers being hotter than with the other species. This can be traced to the molecular cross sections of these species in fig. \ref{fig:cross_secs}. Both MgH and AlO display strong cross sections in the wavelength region near 0.5$\micron$ which drop away significantly nearer to the infrared. AlO displays a lower cross section near 0.6$\micron$ and the cross section of MgH becomes almost completely negligible above 0.8$\micron$. This means that unlike the other metallic species, there are opacity windows in which the stellar irradiation is able to penetrate down to the higher pressures. This in turn causes the deeper layers of the atmosphere to heat up, hence why the temperature of the deep atmosphere is the highest with these two species. This also shifts the thermal inversion to higher in the atmosphere. The other metallic species (TiO, VO, CaO and NaH) display strong cross sections throughout the visible region (<1$\micron$) due to which the deepest layers of these model atmosphere are left relatively cool.

Some equilibrium P-T profiles in fig. \ref{fig:double_plots} also show small non-inverted profiles near $\sim1$ bar before the inversion at P$\lesssim0.1$ bar in the atmosphere, particularly for species such as AlO. This dip in the P-T profile is contributed by the variation in the value of $\gamma \equiv \kappa_\mathrm{vis}/\kappa_\mathrm{ir}$ as a function of pressure. We found that the overall value of $\gamma$ decreases for pressures greater than 1 bar. This is primarily dominated by the broadening of the molecular features, which result in a change in the overall opacity. The broadening of the lighter volatile species with pressure is greater than those of the much more massive metallic species such as TiO and VO. Thus at P$\sim1$ bar the $\kappa_\mathrm{ir}$ is often dominant over the visible opacity. At lower pressures, however, the metallic species dominates the opacity, and thus $\gamma>1$ at higher altitudes, resulting in an inversion at pressures below 0.1 bar. In addition, species such as MgH and AlO have significant opacity windows (see fig. \ref{fig:cross_secs}) where radiation flux can penetrate deeper in the atmosphere (as discussed above) to heat these layers just below the photosphere.

All of these species are capable of causing thermal inversions in hot Jupiters. At solar elemental abundance most of the species are at or near the abundances required to form inversions. It is quite likely that multiple such species may contribute to the visible opacity and thus require lower abundances of each for an inversion. How much we expect of such species in the atmospheres of hot Jupiters is an unknown, given that as well as the chemical equilibrium, disequilibrium effects may act to enhance/diminish the abundance. Another assumption here is that all of the nuclei of the relevant metal are bonded to the given molecule, which may not be the case depending on the temperature and pressure dependant chemical equilibrium of the atmosphere. For instance, the Mg may partially form Magnesium silicate clouds in the atmosphere instead of MgH, thus reducing the overall atmospheric abundance. Mixing from the night side and/or upwelling from deeper in the atmosphere may also alter the abundance. However, what we have been able to demonstrate here is that as well as TiO and VO, other species are also capable of forming thermal inversions on the very hottest extrasolar giant planets. We have also demonstrated the effect of the individual species on the equilibrium P-T profile, and how there is variation with the molecular cross section of each species compared to the simple semi-grey model seen in section \ref{sec:grey}. Our whole focus so far has been on the visible opacity, $\kappa_\mathrm{vis}$, but in section \ref{sec:c/o} we will explore how the infrared opacity from the 7 volatile species can also alter the P-T profile.

\subsection{Inversions on Very Hot Planets}

The temperature on hot Jupiters for these newer species to exist in the gaseous phase is however very high, $\sim$2500K \citep{sharp_2007,burrows_2009}, and only the most strongly irradiated exoplanets are going to have conditions sufficient for them to remain in their gaseous form. Temperatures such as this have been observed for several systems, e.g WASP-33b and WASP-18b. We do observe thermal inversions in both of these systems \citep{haynes_2015, sheppard_2017}. Recent observations of thermal emission from other very hot Jupiters equilibrium temperatures greater than 2000K) has revealed some surprising insights into the atmospheric temperature profile. Many of these extremely hot planets studied with WFC3 data show very few spectral features. This indicates that the temperature may be close to isothermal in these atmospheres. If this is the case it would mean that the infrared opacity does equal the visible opacity. This would be an important observation for hot Jupiter atmospheres as it shows the visible opacity is significant for these exoplanets. Photodissociation of H$_2$O has also been proposed at these high temperatures \citep{parmentier_2018}. This would reduce the strength of spectral features in the infrared due to the loss of H$_2$O. In addition, the formation of H- in the atmosphere may result in strong visible opacity due to its cross section. This would act to increase the strength and possibility of an inversion as it would increase the overall visible to infrared opacity ratio \citep{lothringer2018, parmentier_2018}. \cite{lothringer2018} has proposed various candidate species (e.g. Fe, Mg and SiO) which may provide sufficient visible opacity in order to form thermal inversions in atmospheres of planets such as KELT-9b, with an equilibrium temperature exceeding 4000K.

\subsection{Inversions on Cooler Planets}

Cooler hot Jupiters with equilibrium temperatures between 1000-2000K are less likely to have a thermal inversion from the metallic species considered above. None of the species considered here remain gaseous at these comparatively low temperatures and therefore $\kappa_\mathrm{vis}$ remains lower than $\kappa_\mathrm{ir}$, preventing an inversion from occurring. However, other species have been proposed which exist in the gaseous phase at temperatures below 2000K and which may have strong visible cross sections. \cite{zahnle_2009} showed that at these intermediate temperatures absorption from species such as HS or S$_2$ is sufficient to produce thermal inversions thanks to their strong UV absorption. These species may however photo-dissociate under strong irradiation \citep{knutson_2010}. Further exploration of the disequilibrium chemistry may suggest the feasibility of thermal inversions by these sulphur bearing species. Other proposed species capable of thermal inversions at these intermediate temperatures include atomic Na and K \citep{molliere_2015}.

\section{Effect of Infrared Opacity}
\label{sec:c/o}

\begin{figure}
	\includegraphics[width=\columnwidth]{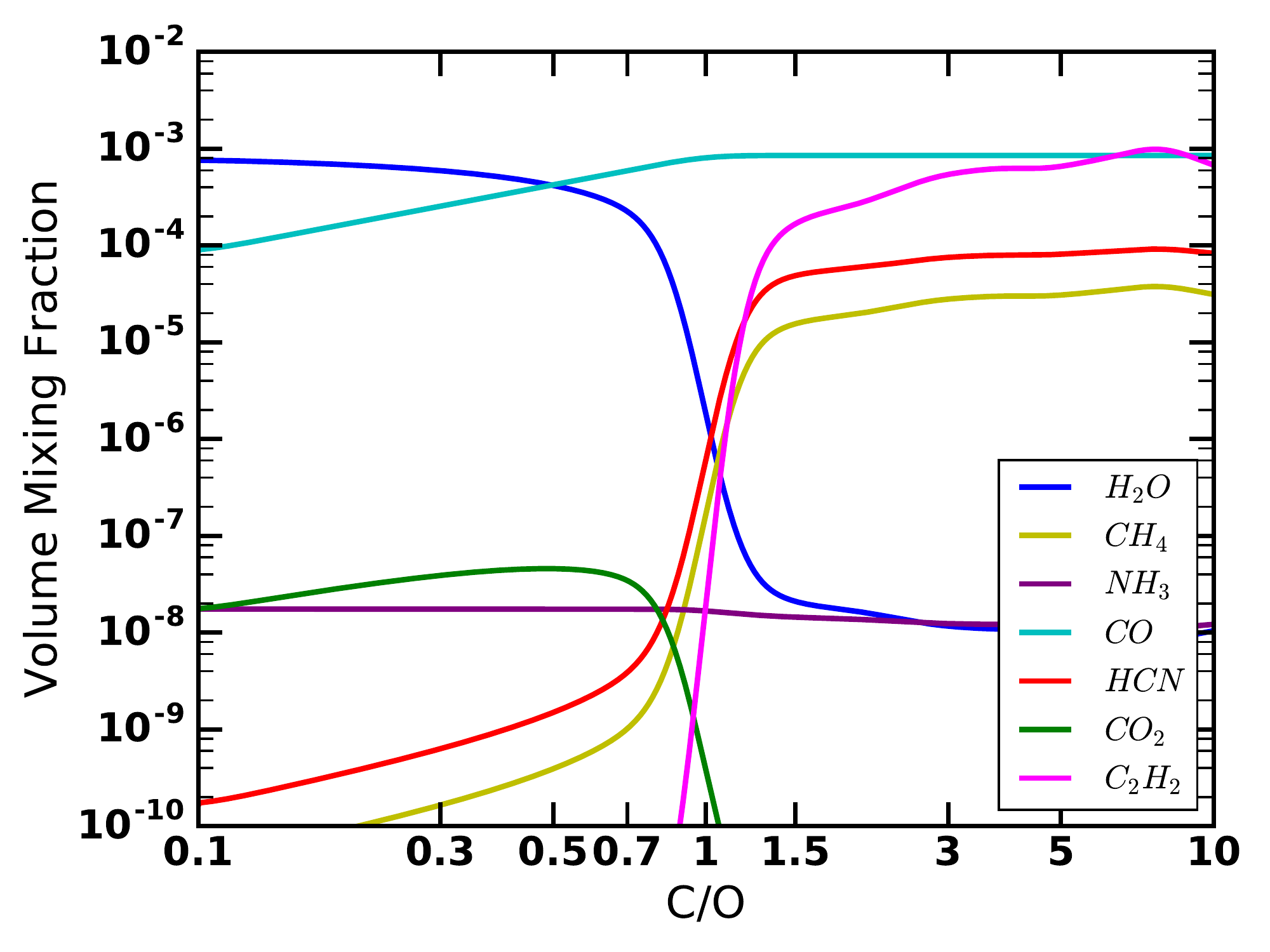}
    \caption{Equilibrium mixing ratios for the 7 volatile species which have significant absorption in the infrared as a function of the C/O ratio. This was calculated at a temperature of 2000K and 0.1 bar pressure.}
    \label{fig:eqm_chem}
\end{figure}

\begin{figure}
	\includegraphics[width=\columnwidth]{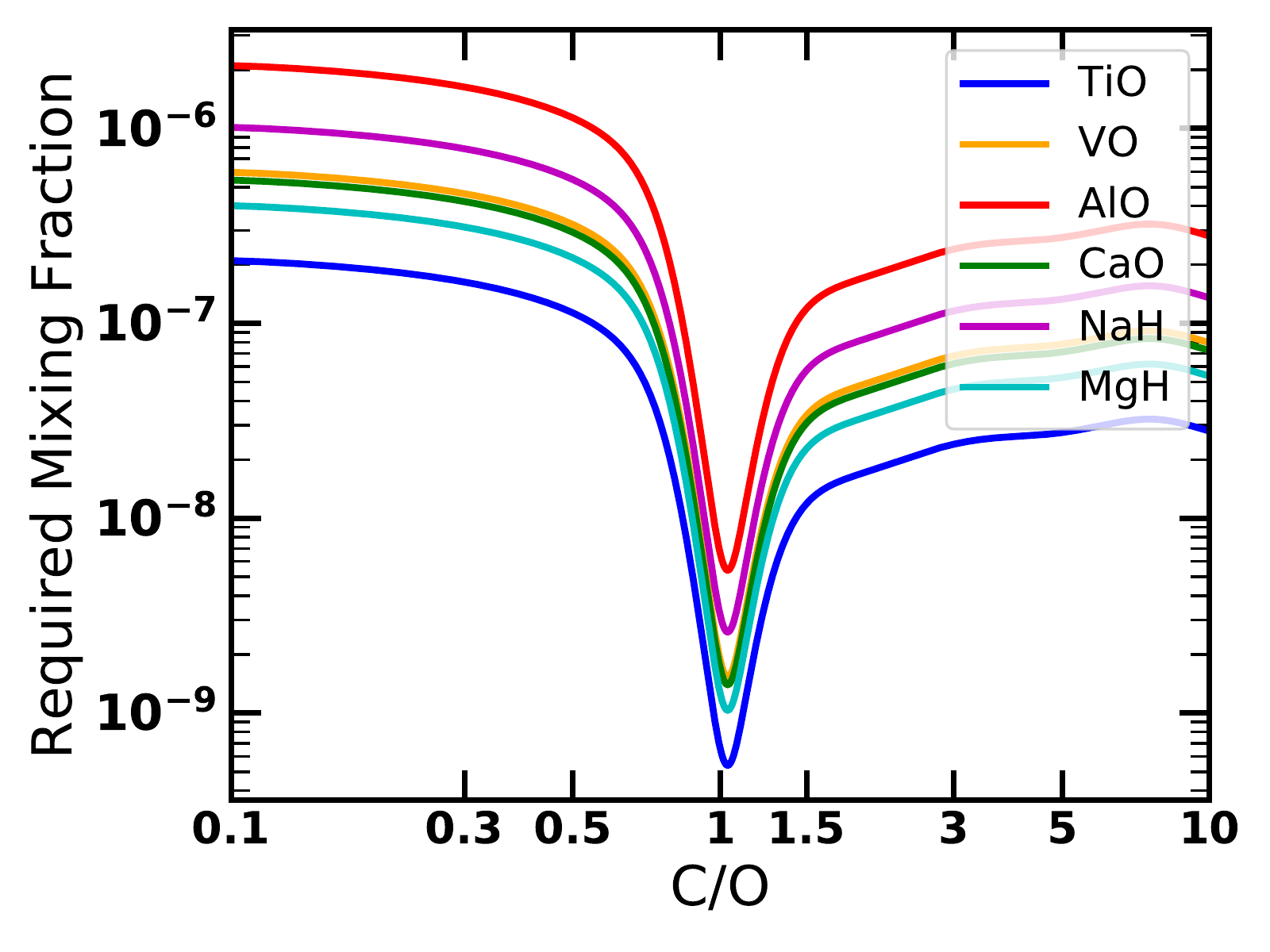}
    \caption{ Required molecular abundance for a thermal inversion for each of the metallic species as a function of the C/O ratio. These were calculated assuming the semi-grey model such that the visible opacity equals the infrared opacity. The model atmosphere was assumed to have an equilibrium temperature of 2000K.}
    \label{fig:inv_abundance}
\end{figure}

Variation of the infrared opacity, $\kappa_\mathrm{ir}$, can also alter $\gamma$ such that we can satisfy the inequality in equation \ref{eqn:gamma}. This is an area that has not been explored in great detail, but the infrared opacity is just as significant in determining the conditions for thermal inversions as the visible. If $\kappa_\mathrm{ir}$ is reduced whilst holding $\kappa_\mathrm{vis}$ fixed, the visible opacity will eventually exceed the infrared opacity. This will lead to thermal inversions in the photosphere without the need for high abundances for species with strong visible cross sections. We will vary the infrared opacity in the atmosphere through the atmospheric C/O ratio. The motivation for this was two-fold. Firstly, we may explore the variation in the infrared opacity using a single parameter and whilst maintaining chemical equilibrium for the volatile molecular species (see fig. \ref{fig:eqm_chem}). This allows us to explore the more complex parameter space of these species and their chemical reactions without having to use unphysical atmospheric parameters. Secondly, the variation of the C/O ratio results in a very large variation in the overall infrared opacity of the atmosphere, especially due to the low H$_2$O abundance at C/O>1. Up to 3 orders of magnitude difference can be achieved by varying the C/O ratio by only a factor of $\sim$2 due to the very sharp change in equilibrium chemistry near C/O$\sim$1 (see fig. \ref{fig:eqm_chem}). Therefore we are able to significantly alter $\gamma$ and thereby lead to inversions with only relatively small changes in the atomic abundances.

\subsection{Chemical Equilibrium of Volatile Species}

The equilibrium chemical abundance of the volatile species can vary by many orders of magnitude as the C/O ratio is altered. Fig. \ref{fig:eqm_chem} shows the equilibrium abundance of H$_2$O, CH$_4$, NH$_3$, CO, CO$_2$, HCN and C$_2$H$_2$ as a function of the C/O ratio at a temperature of 2000K and 0.1 bar pressure following \cite{madhu_2012}. Low C/O ratios such as solar values ($\sim$0.54) are dominated by water vapour and CO. The remaining species are only present in very small quantities, with the exception being N$_2$, not shown in fig. \ref{fig:eqm_chem} as it has no significant spectroscopic signature. Increasing the C/O ratio towards 1 increases the CO abundance linearly until C/O$>$1, where it levels off given that its abundance is now limited by the available oxygen. As there is now a dearth of O, oxygen bearing species such as H$_2$O and CO$_2$ are greatly depleted in the atmosphere. On the other hand, carbon rich species such as CH$_4$ and C$_2$H$_2$ are now enhanced in the atmosphere, as well as HCN. This change in the abundance by several orders of magnitude results in a significant change to the opacity.

The infrared opacity is most strongly influenced by H$_2$O. The cross section for H$_2$O is stronger than the other species and therefore the atmospheric H$_2$O abundance effects the $\kappa_\mathrm{ir}$ the most strongly. Therefore, as the C/O ratio is increased towards 1, the H$_2$O and thus the total infrared opacity decreases as shown in fig. \ref{fig:eqm_chem}. This has previously been explored in the work by \cite{molliere_2015}. At a C/O ratio of 1, the most abundant species is CO. This therefore greatly reduces the $\kappa_\mathrm{vis}$ required for a thermal inversion in the atmosphere, i.e to satisfy equation \ref{eqn:gamma}. With C/O$>$1, the volume mixing ratio of the carbon rich species now increases and therefore $\kappa_\mathrm{ir}$ increases accordingly. We explore the required abundances of the metallic species with a strong visible opacity to cause thermal inversions in the atmosphere as a function of the C/O ratio.

\subsection{Thermal Inversions with Super-Solar C/O Ratio}

As the C/O ratio is increased towards 1, the required visible opacity in order to produce a thermal inversion decreases. As a result, the required abundance for the species with strong visible cross sections for thermally inverted photospheres is reduced as shown in fig. \ref{fig:inv_abundance}. The lower infrared opacity means a lower visible opacity is needed to satisfy equation \ref{eqn:gamma}. As expected, TiO possesses the strongest opacity and therefore the required abundance for an inversion is the least compared to the other species we consider in our study. The 4 new species, AlO, CaO, NaH and MgH all require $\sim$1.5-2 orders of magnitude higher abundance than the TiO due to their weaker opacity. 

What is uncertain however is the abundances of the metallic species as the C/O ratio increases. As oxygen is most likely in the CO, it unclear whether the oxygen bearing species such as TiO, VO, AlO and CaO would be present in sufficient quantities. Equilibrium calculations have shown that with a C/O ratio of 1, the abundances of TiO and VO reduce by a factor of $\sim$100-1000 \citep{madhu_2011}. However, these species would not need to be at solar abundances to lead to inversions when the C/O ratio is 1. Fig. \ref{fig:inv_abundance} and table \ref{tab:metal_inv} show the required molecular mixing fraction of each species in order to invert the P-T profile as a function of the C/O ratio. To match the infrared opacity, these species can be sub-solar and still lead to a thermal inversion. Therefore regardless of the lower abundance TiO and VO may still be capable of leading to thermal inversion on carbon rich planets. With a C/O ratio of 1 even atomic species such as Na/K are capable of producing thermal inversions in the photosphere of hot Jupiters around some stars. A discussion of the Na/K opacity and its effect at C/O$=1$ can be found in \cite{molliere_2015}. The required mixing fraction of all of the metallic species has dropped by a similar quantity thanks to the lower infrared opacity.

\begin{figure*}
	\includegraphics[width=\textwidth]{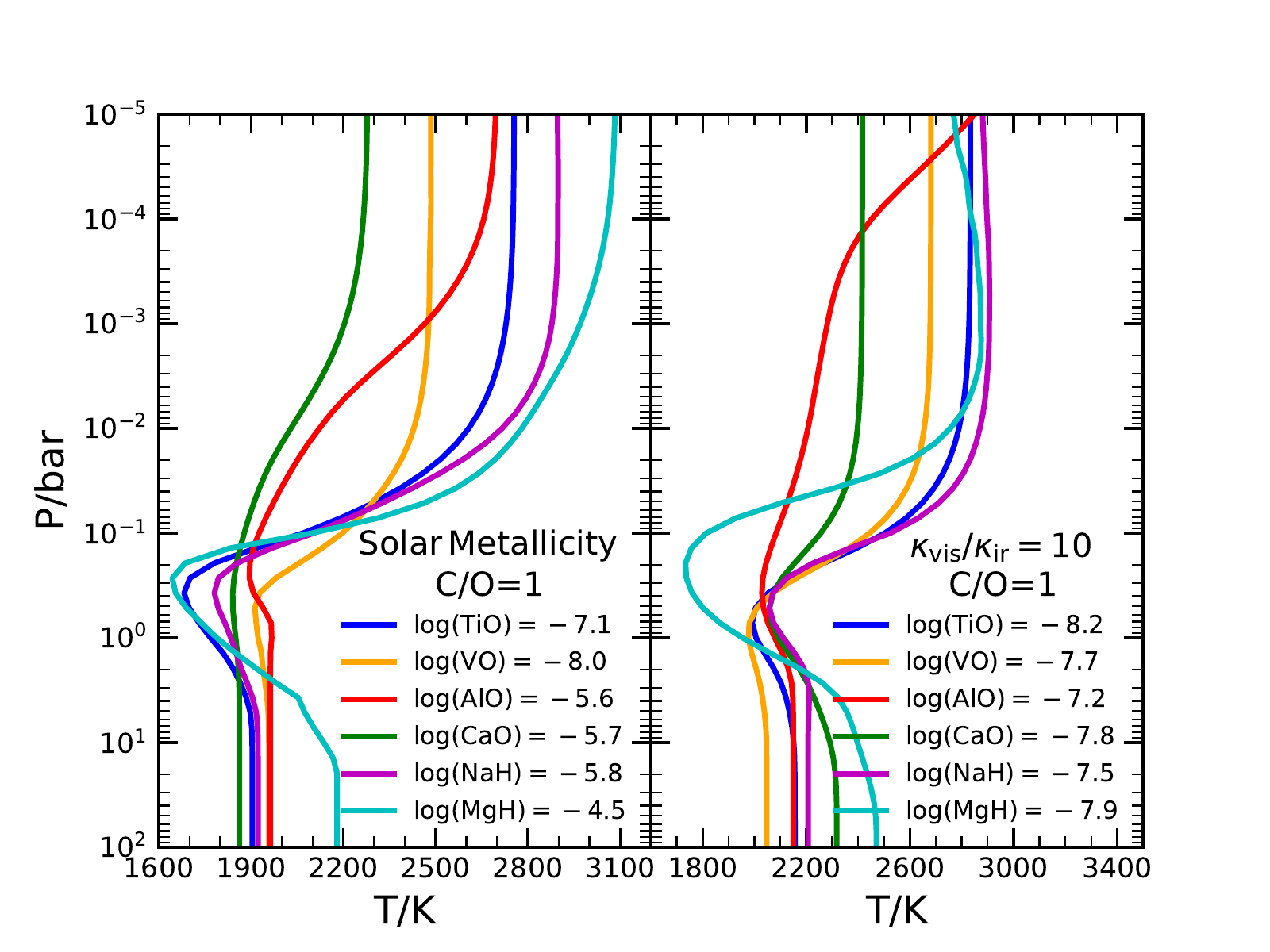}
    \caption{ Radiative equilibrium P-T profiles of model atmospheres with varying abundances of metallic species for C/O = 1. The model assumes a hot Jupiter with an equilibrium temperature of 2000K. The left panel shows the effect of solar mixing fractions of the metallic species, and the right panel shows the abundances such that the ratio of visible to infrared opacity was 10. The volatile species were assumed to be in chemical equilibrium at C/O = 1, and in each model the species was included in the atmosphere by itself with no other visible absorbers.}
    \label{fig:c/o_10}
\end{figure*}

Fig. \ref{fig:c/o_10} shows the effect of solar abundances and $\kappa_\mathrm{vis}/\kappa_\mathrm{ir}=10$ on the P-T profile when the C/O ratio equals 1. The volatile molecular species are taken to be in chemical equilibrium in each layer of the atmosphere. Now, in contrast to fig. \ref{fig:double_plots}, all of the metallic species are abundant enough at solar composition to lead to strong thermal inversions in the photosphere (left side if fig. \ref{fig:c/o_10}). This can be seen from both fig. \ref{fig:inv_abundance} and table \ref{tab:metal_inv}. The visible opacity in the left panel is the same as fig. \ref{fig:double_plots}, but the infrared opacity has now decreased for a C/O ratio of unity. The overall $\kappa_\mathrm{ir}$ has dropped as the H$_2$O, which is the dominant source of opacity in the infrared, is now at a lower mixing fraction. With such little infrared opacity, solar composition of species such as TiO can lead to an inversion with $\Delta T$ of over 1000K. Therefore, should it be the case that high C/O atmospheres still have some visible opacity sources, we may expect that thermal inversions are more likely at C/O$\sim$1, even when the abundances of visible opacity sources are significantly sub-solar. The condition $\kappa_\mathrm{vis}/\kappa_\mathrm{ir}=10$ requires $\sim1$ dex lower mixing fraction for all of the metallic species than solar, such is the weakness of $\kappa_\mathrm{ir}$ when C/O=1. The profiles agree well with each other here, with the temperature in the deeper atmosphere ($> 0.1$bar) very similar for all of the metallic species. Once again though the AlO leads to an inversion at a higher point in the atmosphere than the other species, most likely caused by its lower visible cross section. 

\begin{figure*}
	\includegraphics[width=\textwidth]{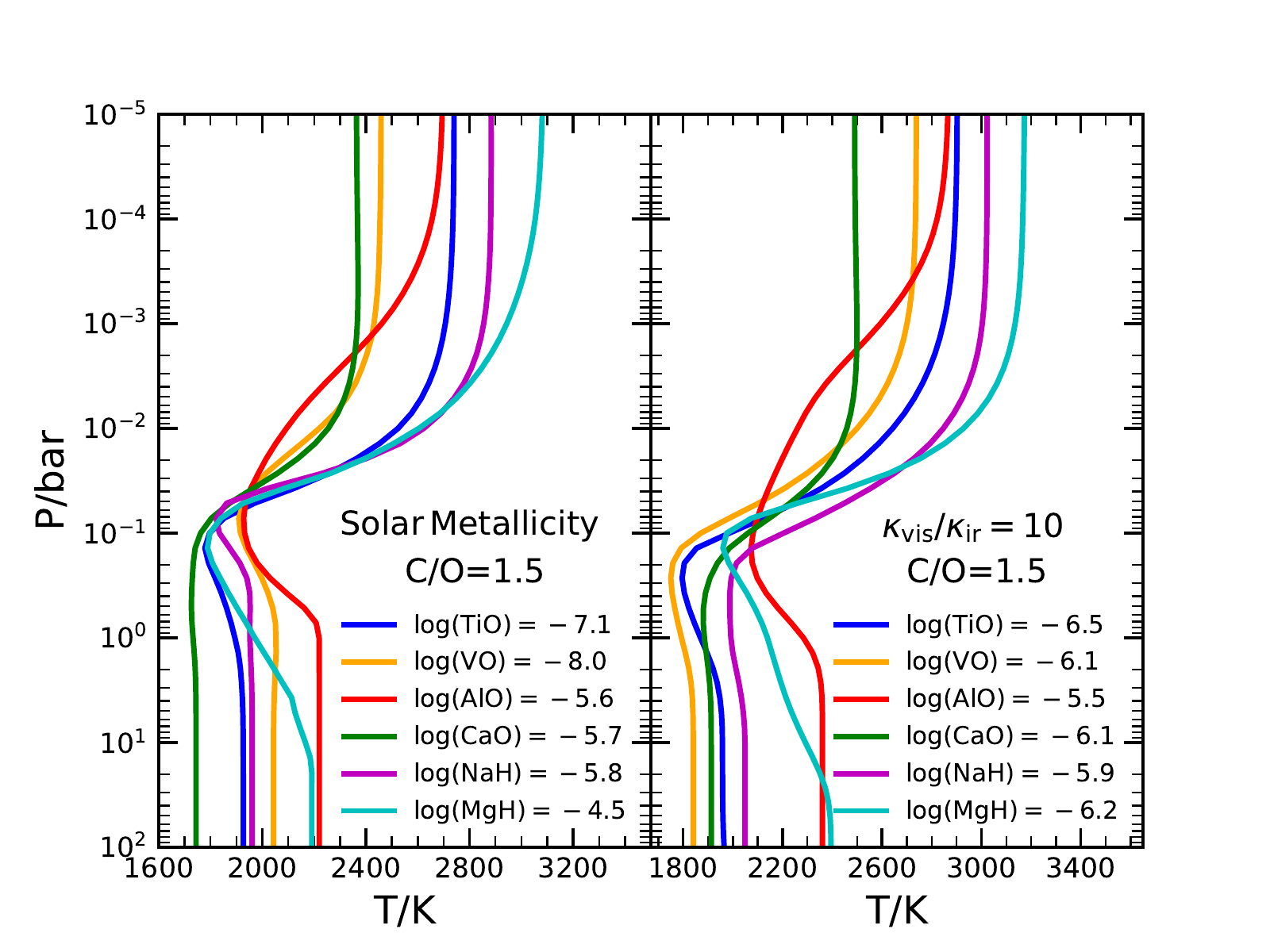}
    \caption{ Radiative equilibrium P-T profiles of model atmospheres with varying abundances of metallic species for C/O = 1.5. The model assumes a hot Jupiter with an equilibrium temperature of 2000K. The left panel shows the effect of solar mixing fractions of the metallic species, and the right panel shows the abundances such that the ratio of visible to infrared opacity was 10. The volatile species were assumed to be in chemical equilibrium at C/O = 1.5, and in each model the species was included in the atmosphere by itself with no other visible absorbers.}
    \label{fig:c/o_15}
\end{figure*}

As the C/O ratio is increased to 1.5 the infrared opacity remains relatively low. Table \ref{tab:metal_inv} and figure \ref{fig:inv_abundance} show that the required abundance for a thermal inversion has increased but is still well below that for a solar C/O ratio. The dominant source of the infrared opacity is now species such as CH$_4$ and C$_2$H$_2$ which can be highly abundant in such carbon-rich atmospheres \citep{madhu_2011}, as shown in fig. \ref{fig:eqm_chem}. All of the species at solar abundance are still capable of leading to thermal inversions in the left hand panel of fig. \ref{fig:c/o_15}. All of the species are within 1 dex of the mixing fraction required to satisfy $\kappa_\mathrm{vis}/\kappa_\mathrm{ir}=10$, therefore both sides of fig. \ref{fig:c/o_15} shows similar P-T profiles. It is interesting to note that despite varying the C/O ratios which leads to $\kappa_\mathrm{ir}$ varying by many orders of magnitude, thermal inversions always form in the photosphere near P$=0.1$ bar. It should also be noted that when the C/O ratio exceeds 1, other larger hydrocarbon molecules not included here may also form in the atmosphere with significant opacity both in the infrared and possibly the visible and thus affect the inversions. With these detailed atmospheric models we will now model a planet with a known thermal inversion, WASP-121b \citep{evans_2017}. 

\section{Case Study: Hot Jupiter WASP-121b}
\label{sec:Real system}

WASP-121b is a hot Jupiter with a radius of $\sim$1.7R$_J$ orbiting an F6 star with an equilibrium temperature of $\sim$2300K. Emission spectroscopy of this planet has shown the presence of a thermal inversion from HST WFC3 observations \citep{evans_2017}. Transmission and emission spectra have also shown evidence for TiO/VO in the atmosphere of this planet \citep{evans_2016,evans_2017}. This makes the exoplanet a good candidate to study thermal inversions and investigate the species responsible. We model WASP-121b for a C/O ratio of 0.5 and 1 with all of the 6 species we have considered thus far that are capable of leading to a thermal inversion. The abundances are set such that the thermal emission spectra most closely resemble the observations.

\begin{figure*}
	\includegraphics[width=\textwidth]{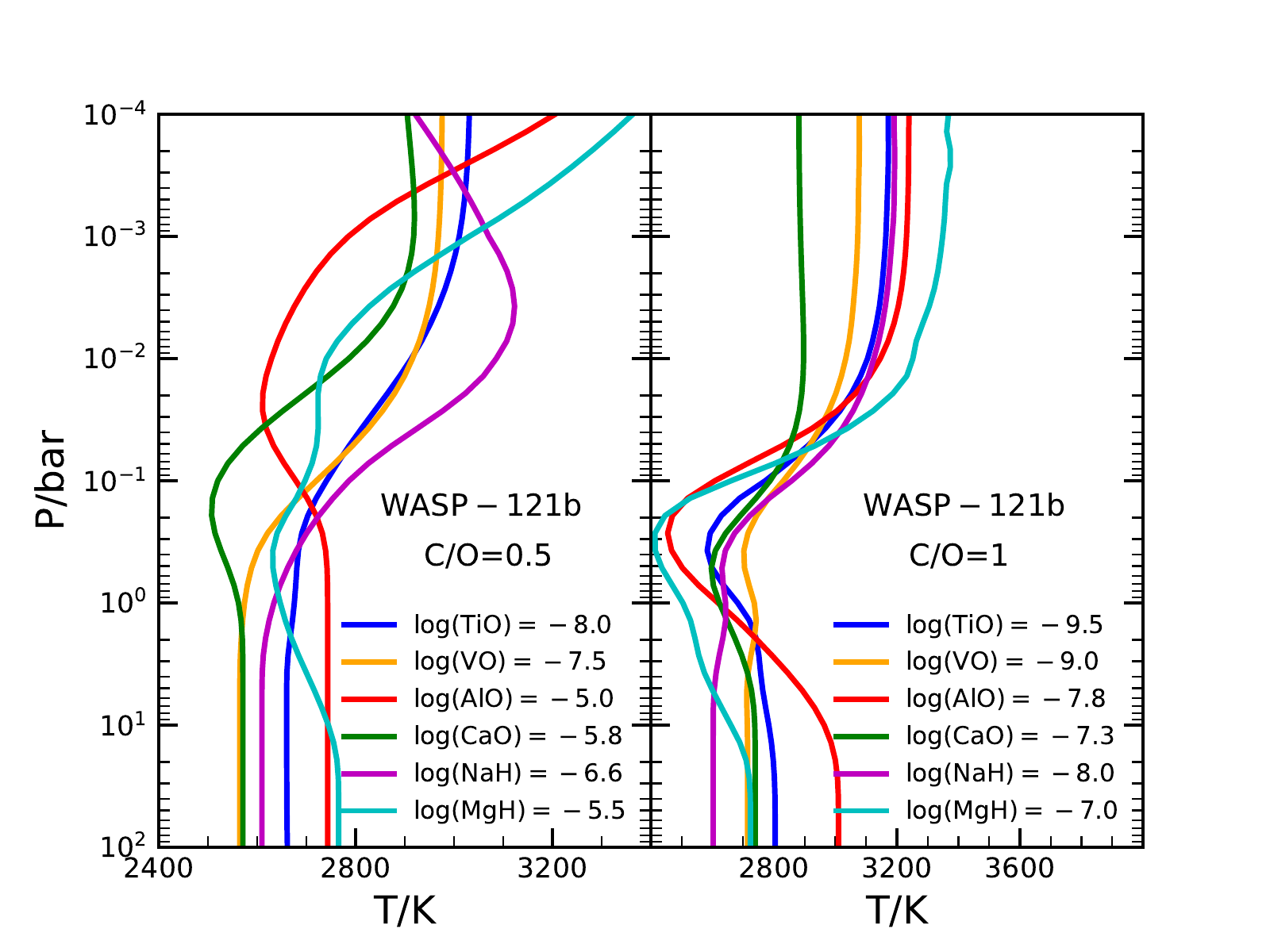}
    \caption{Radiative-convective equilibrium P-T profiles for the planet WASP-121b for a C/O ratio of 0.5 (left) and 1 (right) that most closely match the observed spectrum. In each model the species was included in the atmosphere by itself with no other visible absorbers.}
    \label{fig:wasp121}
\end{figure*}

\begin{figure*}
	\includegraphics[width=\textwidth,trim=4cm 0cm 4cm 0,clip]{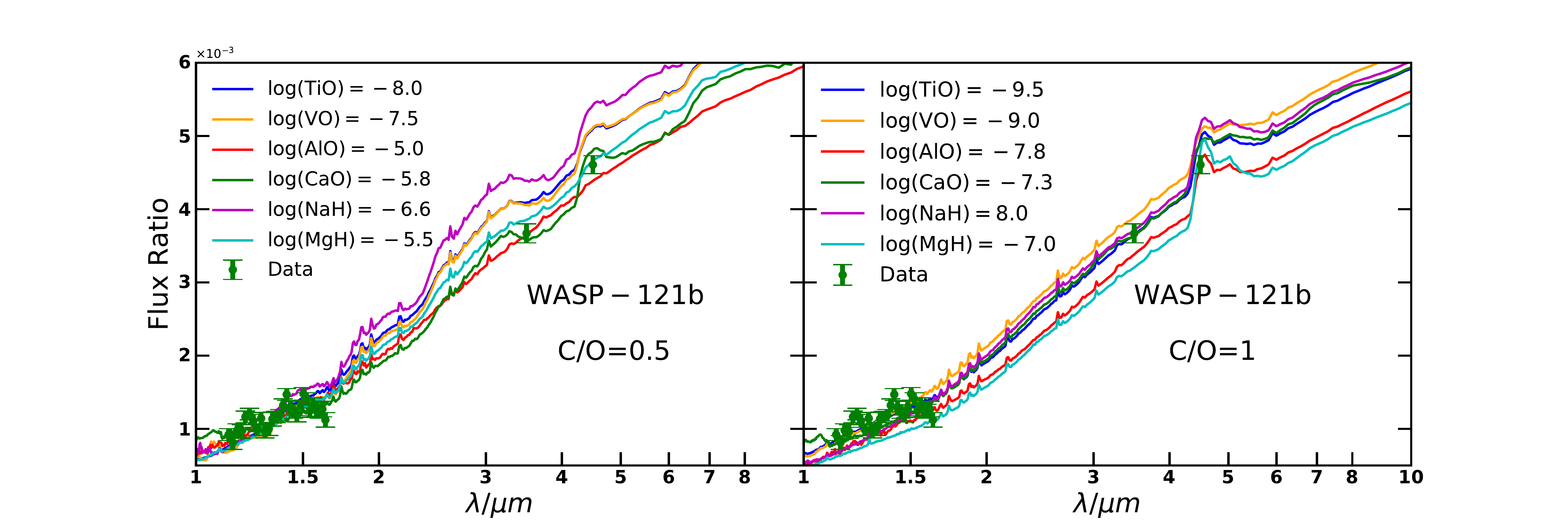}
    \caption{Planet/star flux ratios for the radiative-convective equilibrium models of WASP-121b for a C/O ratio of 0.5 (left) and 1 (right). The green points indicate the Hubble WFC3 and Spitzer photometric observations. The corresponding P-T profiles are shown in fig. \ref{fig:wasp121}.}
    \label{fig:wasp121_flux}
\end{figure*}

Fig. \ref{fig:wasp121} shows the equilibrium P-T profiles for a range of metallic species for each C/O ratio for WASP-121b. The presence of the inversion affects the corresponding spectra in fig. \ref{fig:wasp121_flux} as evident in the emission features. These are clearly visible with all of the species considered, as discussed in section \ref{sec:effect on spectrum}. The required abundances to match the observed spectrum are within $\lesssim$1 dex of solar for all of the species for the C/O=0.5 case (see left panel of fig. \ref{fig:wasp121} and fig. \ref{fig:wasp121_flux}). A C/O ratio of 0.5 also requires a higher abundance of metallic species thanks to the greater infrared opacity due to the H$_2$O content of the atmosphere. As was the case previously, a C/O ratio of 1 reduces the abundances needed for each of the metallic species and therefore sub-solar abundances are sufficient to produce the observed thermal inversion. The required mixing fraction in order to cause an inversion has dropped by $\sim$1.5 dex below that required for a solar C/O ratio. With the C/O ratio of 1, the lack of water vapour or any other strongly IR active species means a lower $\kappa_\mathrm{ir}$ and therefore a lower $\kappa_\mathrm{vis}$ is required for a thermal inversion.

The features of H$_2$O are more prominent on the left panel of fig. \ref{fig:wasp121_flux} and thus provide a stronger spectral feature in the HST WFC3 range compared to the right panel with C/O=1. The spectra follow the observations closer with a C/O ratio of 0.5 than the C/O=1 case, as the H$_2$O emission features are clearly seen in the WFC3 observations. Nevertheless, regardless of the C/O ratio of the atmosphere, we are still able to produce an inversion over both the C/O ratios considered here that matches the observed emission spectrum.

We are able to model the thermal inversion on WASP-121b with 4 new species, AlO, CaO, NaH and MgH along with the well studied TiO and VO. From the GENESIS model we have been able to reproduce the observed emission spectrum for all of the species in radiative-convective equilibrium (fig. \ref{fig:wasp121_flux}). As we have no observations beyond the near infrared, we are unable to further constrain the refractory species responsible for this inversion. Transmission spectroscopy in the optical can further investigate which metallic species are likely to be present in the atmosphere given that they all have strong and unique spectral features below 1$\mu$m.

\section{Discussion and Conclusion}\label{sec:conclusion}

The goal of this work is to characterise the nature of thermal inversions in hot Jupiters and the molecular species which allow thermal inversions to occur. We explored semi-analytic and line-by-line radiative equilibrium models with a number of species thus far not explored and determined 4 new species which are capable of leading to thermal inversions in hot Jupiter atmospheres. We also explored the effect of varying the infrared opacity through the C/O ratio and found that a C/O ratio of 1 results in inversions with the lowest required abundance of metallic species. With this exploration of the parameter space for thermal inversions we compared our findings with WASP-121b, a planet known to have a thermal inversion and which shows evidence for TiO and/or VO \citep{evans_2016,evans_2017}.

As well as the well established TiO and VO \citep{fortney_2008, spiegel_2009}, we found that AlO, CaO, NaH and MgH may also be present in significant enough quantities in order to cause thermal inversions in extrasolar giant planets. Each of these species is capable of causing thermal inversions at or near their solar atomic abundance. We explored the effects of varying their abundance from solar composition \citep{asplund_2009} to $\kappa_\mathrm{vis}/\kappa_\mathrm{ir} = 10$ for a number of different C/O ratios. For each of the metallic species we also obtain the required mixing fraction in order to produce a thermal inversion on a hot Jupiter with an equilibrium temperature of 2000K. These new molecules open up a new avenue to explain thermal inversions in hot Jupiter atmospheres.

Whilst many of these species are capable of causing thermal inversions, there are some caveats and simplifications in our models which we should emphasise. Given their refractory nature, these new species AlO, CaO, NaH and MgH require very high temperatures (>2000K) to be present in the gas phase \citep{sharp_2007, woitke2018}. Therefore, very strongly irradiated atmospheres are required for these species to be present in significant quantities to cause thermal inversions. Such temperatures have been seen for a handful of exoplanets \citep{haynes_2015, cartier_2017, evans_2017, sheppard_2017, beatty_2017_k13}, but many hot Jupiters may be too cool for these species \citep{kreidberg_2014, crouzet_2014, line_2016}. This is also consistent with the fact that thermal inversions have been inferred in only the hottest of hot Jupiters \citep{haynes_2015, evans_2017,sheppard_2017}. Photodissociation of species such as H$_2$O \citep{parmentier_2018} has not been included here, and this may play an important role in determining the overall $\kappa_\mathrm{ir}$ on such high temperature planets. Another important consideration is that we have not explored how equilibrium chemistry affects these metallic species. We assume that the atomic abundance of each of these metallic species is equal to the molecular abundance (i.e. that all of the atoms are bound to the relevant species) but this can be significantly lower for some molecules. Species such as Na have already been observed in hot Jupiters, and thus NaH may not be present at or near its solar atomic abundance in the atmosphere. However, observations of these systems may be able to constrain some of the species to determine how well each is able to cause an inversion.

Optical transmission spectra of such hot Jupiters may provide key insights into the possible metallic species which result in thermal inversions and their atmospheric abundances. Each of these metallic species has a strong opacity at visible wavelengths, with a unique molecular cross section, as shown in fig. \ref{fig:cross_secs}. As such these species are observable with optical transmission spectra. We now have high precision and resolution transmission spectra for a number of planets \citep{nikolov_2014, wyttenbach_2015, chen_2018}, with some even showing evidence for TiO \citep{sedaghati_2017}. With future instruments such as JWST we will be able to obtain even higher precision data down to a wavelength of 0.6$\mu$m, where all of these species possess a strong cross section.

We have also explored variation in the infrared opacity, $\kappa_\mathrm{ir}$, through the C/O ratio. This has been largely ignored until now, but the influence of the infrared active species on thermal inversions is a key step towards understanding the physical processes and emergent spectra. The change in the C/O ratio can result in significant changes in the atmospheric abundance of the volatile species \citep[e.g.][]{madhu_2012, moses_2013} and thus $\kappa_\mathrm{ir}$. We found that the required mixing fraction for a thermal inversion for all of the relevant metallic species decreased as we increased the C/O ratio. The lowest infrared opacity and thus the lowest required mixing fraction of a metal species for a thermal inversion was when the C/O ratio was close to unity. At these C/O ratios, the infrared opacity is dominated at such high temperatures by CO, which has only a small molecular cross section. The H$_2$O abundance is low when C/O=1 ($\mathrm{log(H_2O)}<10^{-5}$), and thus there is no significant opacity from water vapour, which possesses the dominant molecular cross section of all of the infrared volatile species. This means that in order to satisfy $\kappa_\mathrm{vis}>\kappa_\mathrm{ir}$ we require a lower visible opacity compared to the C/O = 0.5 case. 

The C/O ratio is likely to strongly affect the abundance of the metallic species, and we should expect the atmosphere to be greatly depleted particularly of oxygen rich species such as TiO, VO, AlO and CaO. For instance, TiO and VO have their abundance reduced by between 2-3 orders of magnitude when the C/O ratio is 1 compared to their solar abundance \citep{madhu_2011}. However, the H$_2$O abundance and thus the overall infrared opacity falls by a similar amount. Therefore  the required molecular abundance for an inversion-causing metallic species also drops by over 2 orders of magnitude from its solar abundance. A log mixing fraction as low as -10 for species such as TiO is sufficient to satisfy equation \ref{eqn:gamma} and as shown in fig. \ref{fig:inv_abundance}. This means that atmospheres which have a C/O$\approx 1$ may be just as likely to have a thermal inversion even if the visible absorbers are present in significantly sub-solar quantities. This also means that weaker visible absorbers such as Na and K can also affect the inversion \citep{molliere_2015}.

We modelled the thermal inversions on WASP-121b \citep[][]{evans_2017}, a hot Jupiter observed to have a thermal inversion in the retrieved emission spectrum. We modelled all of the metallic species that are capable of leading to thermal inversions and found the abundances that are required to produce a thermal inversion for both C/O=0.5 and C/O=1. We found that we are able to match the observed emission spectrum with every species over both C/O ratios, as shown in figs. \ref{fig:wasp121} and \ref{fig:wasp121_flux}.

Inversions on planetary systems throughout the solar system are common despite the compositional range of the planets that are present, and strongly irradiated hot Jupiters are ideal laboratories to study such inversions. They have strong spectral signatures, and even with current facilities we can obtain high resolution and high precision spectra from their dayside. Studying how and why these planets possess thermal inversions can not only tell us about the landscape of the hot Jupiter systems and their chemistry, but also provide key insights into the formation mechanisms of such planets. This is particularly so given the strong effect the C/O ratio has on the overall infrared opacity and thus the likelihood for an inversion. With current and future instrumentation such as the VLT and JWST we can probe in even more detail the chemical structure and temperature profile of such atmospheres and learn more about the processes that occur on such systems.

\section*{Acknowledgements}

SG and NM acknowledge financial support from the Science and Technology Facilities Council (STFC), UK. We thank Ivan Hubeny for the very valuable comments on the manuscript.



\bibliographystyle{mnras}
\bibliography{references} 







\bsp	
\label{lastpage}
\end{document}